\documentclass{IEEEcsmag}

\usepackage[colorlinks,urlcolor=blue,linkcolor=blue,citecolor=blue]{hyperref}
\expandafter\def\expandafter\UrlBreaks\expandafter{\UrlBreaks\do\/\do\*\do\-\do\~\do\'\do\"\do\-}
\usepackage{upmath,color}

\usepackage{graphicx}
\usepackage{subfigure}

\usepackage{url}
\usepackage{hyperref}
\usepackage{booktabs}
\usepackage{amsmath}

\begin{document}

\title{Deepfake in the Metaverse: An Outlook Survey}

\author{Haojie Wu}
\affil{University of Science and Technology of China, China}

\author{Pan Hui,~\IEEEmembership{Fellow,~IEEE}}
\affil{The Hong Kong University of Science and Technology (Guangzhou), China}

\author{Pengyuan Zhou,~\IEEEmembership{Member,~IEEE}}
\affil{University of Science and Technology of China, China}





\begin{abstract}\looseness-1We envision deepfake technologies, which synthesize realistic fake images and videos, will play an important role in the future metaverse. While enhancing users' immersion and experience with synthesized virtual characters and scenes, deepfake can cause  serious consequences if used for fraud, impersonation, and dissemination of fake information. In this paper, we introduce the principles, applications, and risks of deepfake technology, and propose some countermeasures to help users and developers in the metaverse deal with the challenges brought by deepfake technologies. Further, we provide an outlook on the future development of deepfake in the metaverse.

\end{abstract}

\maketitle

\chapteri{T}he term ``deepfake'' is derived from the combination of ``deep learning'' and ``fake'', initially used by a Reddit user who published fictitious movies on the social media platform in December 2017 that mapped the faces of stars like Scarlett Johansson onto sexual performers~\cite{2017deepfake}. There is currently no internationally accepted definition of ``deepfake''. The ``Malicious Deep Fake Prohibition Act of 2018''~\cite{S.3805} in the United States defines a ``deep fake'' as ``an audiovisual record created or altered in such a manner that the record would falsely appear to a reasonable observer to be an authentic recording of an actual person's speech or conduct'', where ``audiovisual record'' refers to digital content such as images, videos, and voice recordings. Deepfakes~\cite{tolosana2020deepfakes} can be used for various creative purposes, such as entertainment, education, and art, but can also be used for illegal activities, such as spreading rumors, defamation, or deception.

The metaverse has the potential to alter how people live, work, learn, and entertain as well as advance human society by allowing people to transcend temporal and spatial boundaries and experience imaginative scenarios, roles, and activities~\cite{lee2021all}. However, the metaverse may also bring issues such as disconnection from reality, invasions of privacy, threats to network security, and, a \textbf{next-gen deepfake era}. 

Unlike deepfake in current cyberspace, deepfake in the metaverse will mostly be based on 3D technology, used to modify or synthesize virtual content in XR, creating seemingly realistic fake content. Deepfake technology in the metaverse can provide users with a more realistic and immersive VR experience. However, it also leaves malicious users with more opportunities to create various fake virtual contents which are more difficult to identify and distinguish in immersive virtual environments.

This prompts us to write this survey from a novel perspective, systematically reviewing and summarizing potential deepfake trends in the metaverse. 
%

 The organization of the rest of this survey is as follows: in \textit{FUNDAMENTALS}, we summarize the basics of deepfake and the metaverse. In \textit{DEEPFAKE IN THE METAVERSE}, we describe the technical implementation and application scenarios of deepfake in the metaverse. Then, we list potential security and privacy issues related to deepfake in the metaverse in \textit{SECURITY AND PRIVACY}, and in Section \textit{FUTURE OUTLOOK}, we discuss the future development direction of deepfake in the metaverse. Finally, we provide our conclusion in \textit{CONCLUSION}.

\section{FUNDAMENTALS}
\label{sec:two}

\subsection{Overview of Deepfake}
Deepfake technology can be classified into two main categories: manipulation of existing multimedia data (such as images, videos, or audio), and synthesis of new multimedia data using deep learning techniques. There are various principles and methods behind deepfake technology, such as Generative Adversarial Networks (GANs)~\cite{yadav2019deepfake}, encoder-decoder networks, Convolutional Neural Networks (CNNs), image transformation networks (Pix2Pix, CycleGAN), Recurrent Neural Networks (RNNs), and so on.


Deepfake has many valuable application scenarios, such as entertainment content production, audience engagement enhancement, and diagnostic image generation. However, deepfake also brings many potential risks and challenges. First, deepfake technology lacks unified standards and regulations on responsibility and punishment and has controversies in copyright, portrait rights, and privacy, etc. Secondly, deepfake poses risks such as trust crisis, social division, psychological effects, etc., affecting people's judgment of true information. Additionally, there is a trend of using deep learning techniques to generate fake biometric signals (such as face images and videos) for biometric spoofing or presentation attacks, posing a more severe threat than fake images and videos to social security and personal privacy.

Therefore, detecting and defending against deepfake content has become one of the hottest research topics. The key to determining detection performance lies in how to choose relevant features that can effectively distinguish between real and fake content, and how to establish models with good classification performance. The following diagram shows the specific process of detection algorithms. 

\begin{figure}[ht]
    \centering
    \includegraphics[width=7cm]{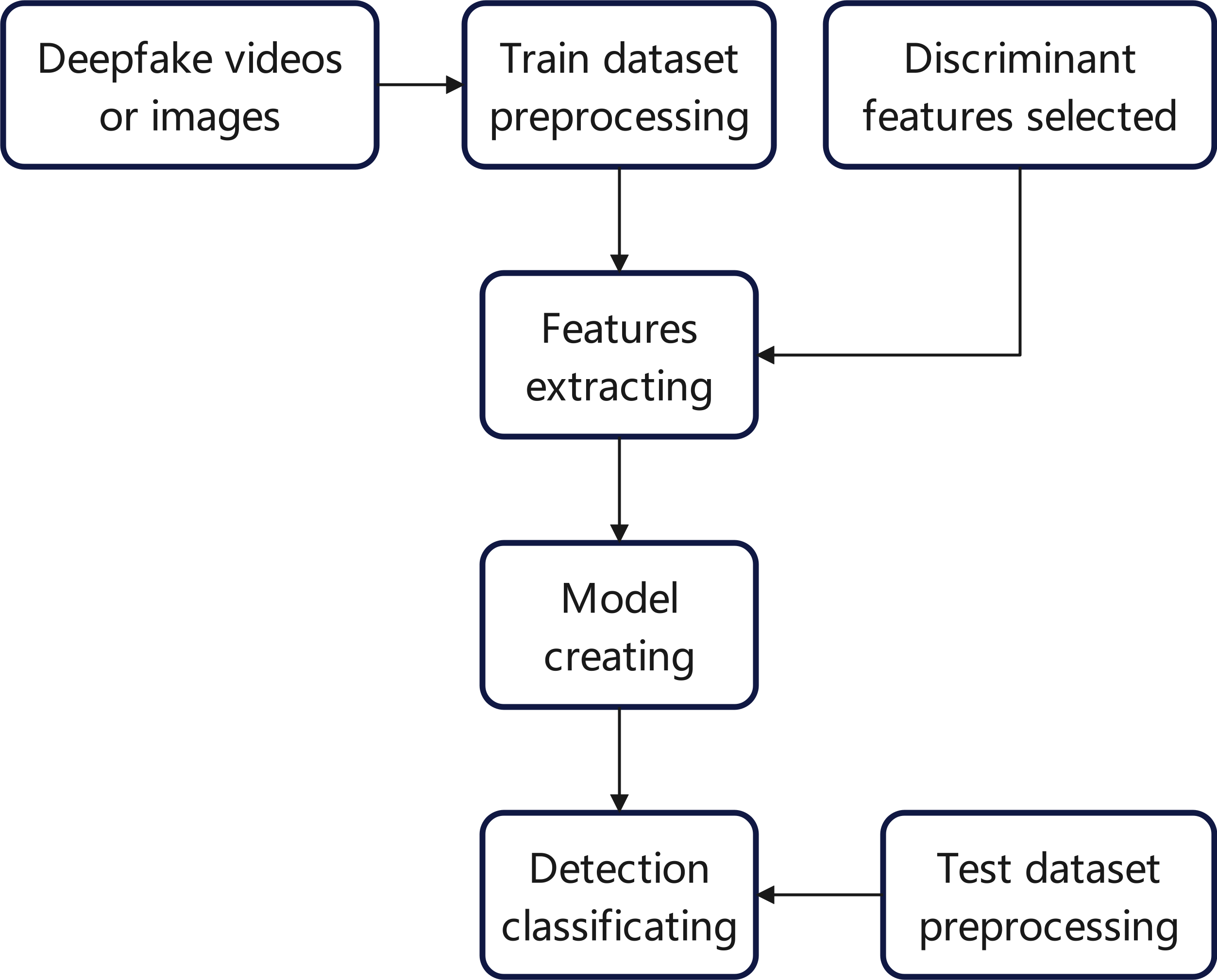}
    \caption{General flowchart of deepfake detection}
    \label{f1}
\end{figure}

Currently, most deepfake works focus on face forgery~\cite{tolosana2020deepfakes}, including facial editing (manipulating individual facial features), facial generation (creating new faces), facial swapping (exchanging one person's face with another), and voice synthesis (recreating sounds) to falsify or modify face images, videos, or audio. 
Face forgery detection~\cite{tolosana2020deepfakes,zhang2022deepfake} can be divided into face forgery image detection and video detection. Different forgery detection methods have different focuses including detection based on specific artifacts, information inconsistency, or data-driven approaches. 

\subsection{Overview of Metaverse}
The metaverse is a virtual parallel world that maps or transcends the real world and interacts with it. It closely integrates the virtual world with the real world in terms of economic systems, social systems, and identity systems, and allows each user to produce content in the virtual-physical blended world. It's widely applicable in social, gaming, education, business, entertainment, advertising, culture, travel, health, etc. For instance, users can participate in online education and training, participate in virtual art exhibitions, buy virtual goods or digital collections, and enjoy virtual trips, among others.

The metaverse is primarily divided into two types: virtual reality (VR) and augmented reality (AR). VR offers an artificial reality that enables body parts (such as hands) to interact with the environment. AR adds a virtual layer on top of the physical world. VR/AR apps require extensive data collection to continuously monitor the user and surrounding for better context understanding, a key factor for immersive experience, and thus face serious risks like data breaches~\cite{4falchuk2018social,5liu2018cooperative}. 

There is a huge overlap between the future development of deepfake and metaverse. On the one hand, deepfake technology can provide richer content and more realistic experiences for the development of the metaverse. On the other hand, deepfake can generate fake content or environments to trick users, which is harder to be noticed due to the immersive experience. 
Therefore, we need to integrate careful design and countermeasures as well as enforce thorough regulations when using deepfake to augment metaverse experience.  

\subsection{Related survey}

Being at its infancy, deepfake in the metaverse has quite limited relevant studies. As shown in Table~\ref{t1}, Tariq et al.~\cite{2023deepfake} examined the security implications of deepfake in the metaverse and proposed potential solutions to mitigate risks in virtual gaming, online meetings, and virtual offices. Vasist et al.~\cite{vasist2022deepfakes} provided an integrative overview of existing deepfake research, highlighting gaps in knowledge and proposing future research directions, aiming to consolidate and map the current state of the field. Mirsky et al.~\cite{mirsky2021creation} provided an in-depth exploration of deepfakes, covering their creation, detection, current trends, defense shortcomings, and areas requiring further research.
Some related surveys~\cite{7wang2022survey,8fernandez2022life,9chen2022metaverse,10zhao2022metaverse} introduced some technologies and application scenarios related to the metaverse, as well as potential security and privacy issues that may arise during its development, or propose some future developments in metaverse security. 

\begin{table*}[h!t]
\scriptsize
\centering
\caption{Related surveys}
\label{t1}
\begin{tabular}{ccc}\toprule
\textbf{Years.}&     \textbf{Refs.}&            \textbf{Contributions.}     \\ \hline

2023 & \cite{2023deepfake}&   \multicolumn{1}{m{11cm}}{Examined the security implications of deepfakes in the metaverse, primarily in the context of gaming, online meetings, and virtual offices, and proposed potential solutions.}\\\hline

2022 & \cite{vasist2022deepfakes}&   \multicolumn{1}{m{11cm}}{Provided an integrative overview of existing deepfake research, highlighting gaps in knowledge and proposing future research directions.}\\\hline

2021 & \cite{mirsky2021creation}& \multicolumn{1}{m{11cm}}{Provided an in-depth exploration of deepfakes, covering their creation, detection, current trends, defense shortcomings, and areas requiring further research.}\\\hline


2022 & \cite{7wang2022survey}&    \multicolumn{1}{m{11cm}}{Introduced the fundamentals, security, and privacy of Metaverse, and discussed the associated security and privacy threats as well as solutions.}\\\hline

2022 & \cite{8fernandez2022life}&   \multicolumn{1}{m{11cm}}{Provided an overview of the challenges and proposes an ethical design framework for the development of the metaverse, focusing on privacy, governance, and ethics.}\\\hline

2022 & \cite{9chen2022metaverse}&   \multicolumn{1}{m{11cm}}{Provided an in-depth examination of the security and privacy issues in Metaverse technologies, offering insights into risks, solutions, and future research directions.}\\\hline

2022 & \cite{10zhao2022metaverse}&   \multicolumn{1}{m{11cm}}{Examined the metaverse concept, its potential benefits, security, and privacy concerns, proposed solutions, and the importance of a philosophical approach for progress.}\\\hline





\textbf{Now } & \textbf{Ours}  &   \multicolumn{1}{m{11cm}}{Focusing on the field of deepfake in the metaverse and presenting a future outlook for deepfake in the metaverse.}\\

\bottomrule
\end{tabular}
\end{table*}

\section{DEEPFAKE IN THE METAVERSE}
\label{sec:three}

Deepfake technology in the metaverse can be used to create interactive virtual characters and scenes in the metaverse, e.g., real-life-looking neighbors for online virtual social networks. Furthermore, users can use deepfake to create virtual content such as movies, games, and artworks in the metaverse.

\subsection{2D Deepfakes VS 3D Deepfakes}

Deepfake technology in the metaverse can be implemented using either 2D or 3D techniques. Most 2D deepfake~\cite{tolosana2020deepfakes} works focus on using deep learning technology in flat images or videos, primarily using facial recognition and replacement on a 2D plane, to synthesize a person's facial expressions, movements, and voice onto another person's body, creating seemingly realistic fake images or videos. This technology can be used to create virtual characters, celebrities, news, etc.


3D deepfake technology is a deep forgery technique based on 3D modeling and rendering~\cite{20203dfacegan}. This technology can reconstruct a 3D controllable head model based on a face image or create a completely fictitious character model, and then process the model through texture mapping, lighting, and animation, generating a realistic 3D video or animation. Compared to 2D deepfake, 3D deepfake technology produce more realistic virtual characters at the cost of higher computation complexity. 

Table~\ref{t_DF} presents the comparisons between 2D deepfake and 3D deepfake. Although 2D and 3D deepfake technologies do share some features, they have their own advantages and disadvantages as well as different application scenarios. The developer can choose to use either or combine the two according to specific demands. For example, developers can use 2D deepfake technology to create fake paintings and place them into the 3D virtual scenes created by 3D deepfake, to allow the users to walk around the paintings.

\begin{table*}[h!t]
\scriptsize
\centering
\caption{Comparison between 2D \& 3D deepfake}
\label{t_DF}
\begin{tabular}{ccc}\toprule
\textbf{   }&     \textbf{2D deepfake.}&            \textbf{3D deepfake.}     \\ \hline

\textbf{Technology}    & \multicolumn{1}{m{4.5cm}}{Image generation technology.}       & \multicolumn{1}{m{7.5cm}}{3D modeling and rendering technology, \textcolor{red}{more complex}.}     \\\hline

\textbf{Realism}      & \multicolumn{1}{m{4.5cm}}{Less to no 3D effect of real objects.}      & \multicolumn{1}{m{7.5cm}}{Realistic and convincing 3D vision, \textcolor{red}{more complex}.}      \\\hline

\textbf{Detection}    & \multicolumn{1}{m{4.5cm}}{Pixel and feature analysis.}    & \multicolumn{1}{m{7.5cm}}{3D modeling analysis, \textcolor{red}{more complex}.}     \\\hline

\textbf{Application}    & \multicolumn{1}{m{4.5cm}}{Face swapping and image manipulations.}      &  \multicolumn{1}{m{7.5cm}}{Creating 3D virtual objects, characters or scenes, \textcolor{red}{more possibilities}.}   \\

\bottomrule
\end{tabular}
\end{table*}

\subsection{Applications}

Deepfake technology combined with AR/VR enables users to create hyper-realistic forged reality, e.g.,  creating or imitating virtual characters and scenes and interacting with other users, enhancing the immersion and creativity of the metaverse. Table~\ref{t_app} lists some application scenarios.


\begin{table*}[h!t]
\scriptsize
\centering
\caption{Deepfake application scenarios in the metaverse}
\label{t_app}
\begin{tabular}{cc}\toprule
\textbf{Application fields.}&     \textbf{Example scenarios.}          \\ \hline

Audio and graphic production & \multicolumn{1}{m{11cm}}{To create content for movies to revive a deceased star or historical figure, such as ``Qian Xuesen Digital Human'' in Figure \ref{f3.1} \footnotemark{}.}\\\hline

Human-computer interaction &    \multicolumn{1}{m{11cm}}{To generate realistic virtual assistants or chatbots, or allow users to experience different identities or environments in VR/AR.}\\\hline

Satire field &   \multicolumn{1}{m{11cm}}{To mimic the words and actions of political figures or celebrities, or make them say ridiculous or funny things, as shown in Figure \ref{f3.2} \footnotemark{}.}\\\hline

Video conferencing &   \multicolumn{1}{m{11cm}}{To allow participants to conduct meetings in different backgrounds or scenes.}\\\hline

Facial recognition &   \multicolumn{1}{m{11cm}}{To allow users to use different faces on different platforms.}\\\hline

Education &   \multicolumn{1}{m{11cm}}{To create an immersive museum experience to enhance visitors' understanding of the museum, as shown in Figure \ref{f3.3} \footnotemark{}.}\\\hline

Healthcare &  \multicolumn{1}{m{11cm}}{To allow doctors to perform surgery or consultations in remote or virtual environments, as shown in Figure \ref{f3.4} \footnotemark{}.}\\\hline

Fashion industry &   \multicolumn{1}{m{11cm}}{To try on virtual clothes at home, making shopping more convenient and intuitive via VR glasses.}\\

\bottomrule
\end{tabular}
\end{table*}

\begin{figure}[ht]
    \centering  

    \subfigure[]{
    \label{f3.1}
    \includegraphics[width=0.23\textwidth]{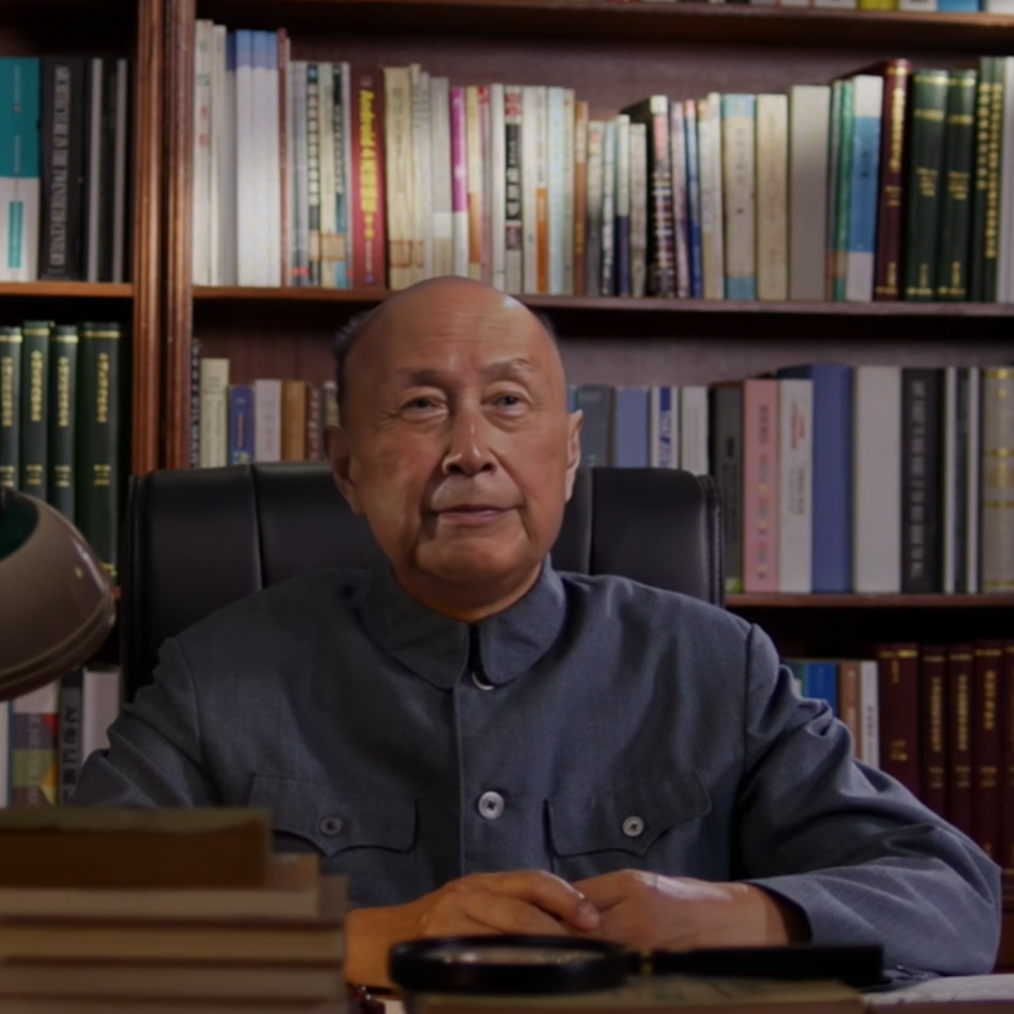}}
    \subfigure[]{
    \label{f3.2}
    \includegraphics[width=0.23\textwidth]{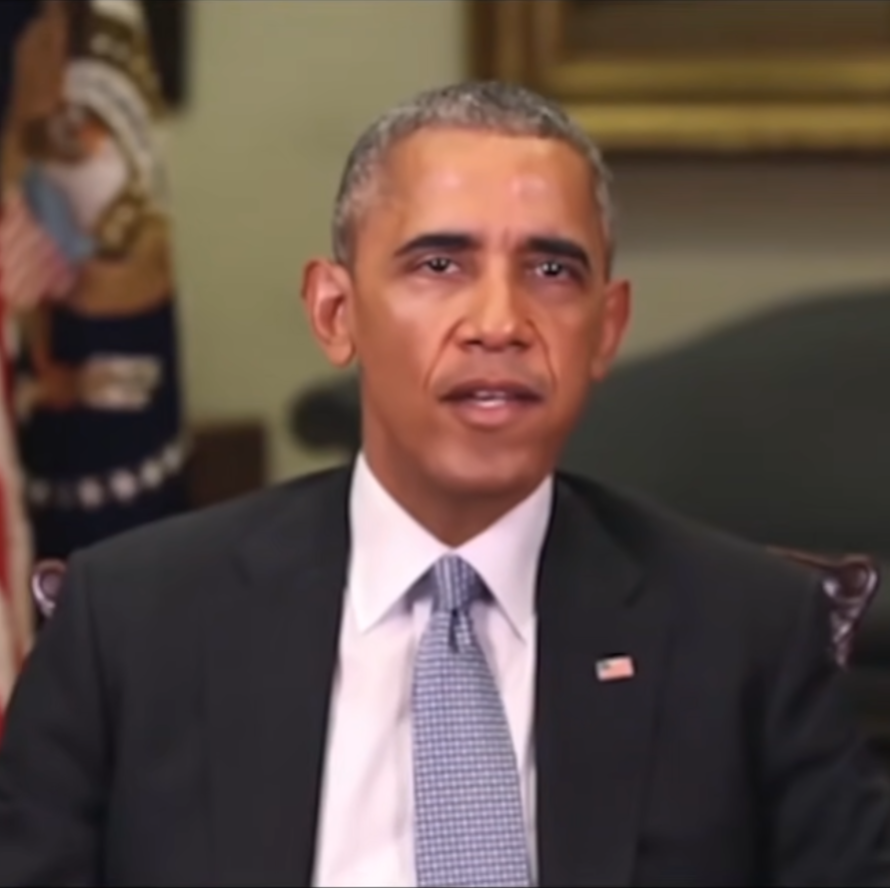}}
    \subfigure[]{
    \label{f3.3}
    \includegraphics[width=0.23\textwidth]{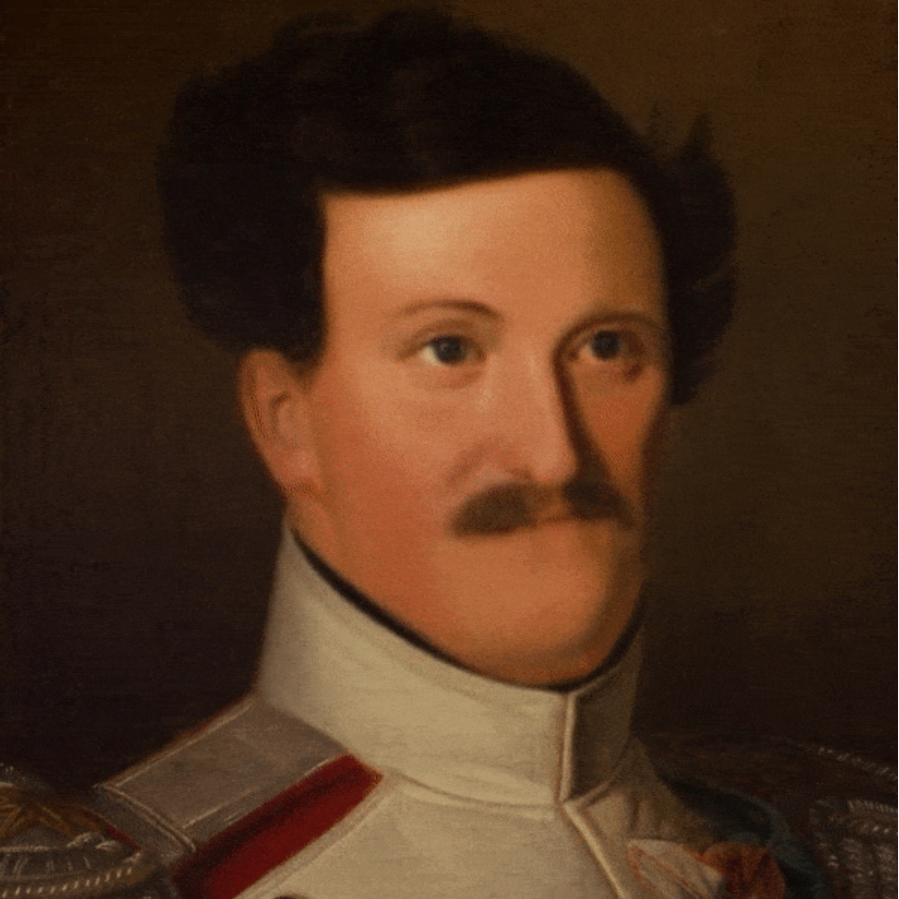}}
    \subfigure[]{
    \label{f3.4}
    \includegraphics[width=0.23\textwidth]{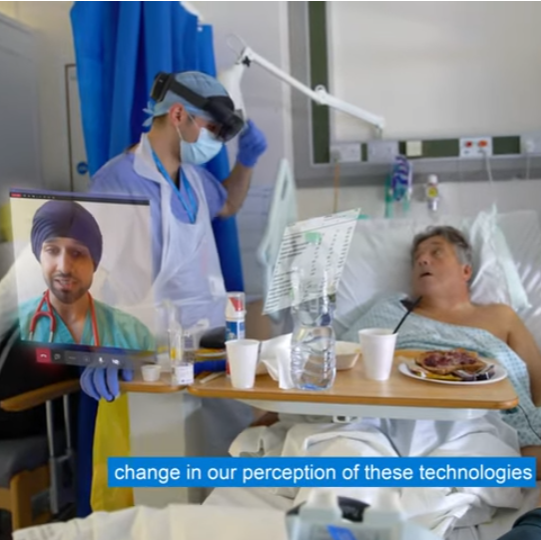}}
    \caption{Examples of deepfake usecases in the metaverse. Refer to Table~\ref{t_app} for details.}
    \label{f3}
\end{figure}



Besides technical implementation and application scenarios, we also elaborate on security and privacy issues that deepfake brings to the development of the metaverse and present some existing countermeasures in the following section. 

\section{SECURITY AND PRIVACY}
\label{sec:four}

In the metaverse, users can freely create personalized faces (can be based on their real faces) using face swapping/deepfake technology, have their own virtual identities, and achieve various immersive experiences. However, it also brings serious security and privacy issues, leading to deception, defaming, or threatening users, companies, or organizations in the metaverse.

\subsection{Security Threats}

The metaverse is expected to collect and analyze personal attributes closely related to users' physiology and behavior, including biometric information such as fingerprints, irises, heart rates, and brainwaves. At the same time, users' interactive behaviors, such as economic transactions and social interactions, will also generate a large amount of valuable and sensitive data. Once this private data is leaked or abused, serious consequences can occur.
For instance, the malicious user can steal a legitimate user's identity and impersonate it to deceive the victim's friends or social network connections, making them believe that they are interacting with the victim, and conduct Internet fraud~\cite{MetaDeepFake,MetaverseDeepFake}. This can be easily achieved in the metaverse because users are used to interacting via digital avatars in the first place. Table~\ref{t_threats} provides a brief summary of the deepfake threats inside and outside the metaverse. Bose and Aarabi~\cite{15bose2019virtual} have shown that deepfake technology can be used in VR to replace a user's face with another one. Using this technology, attackers can impersonate benign users to make financial deals with other users or spread fake virtual information, intentionally causing financial or reputational damage. The following explains in detail the representative security threats brought by deepfake to the metaverse.

\begin{table*}[h!t]
\scriptsize
\centering
\caption{Summary of the deepfake threats inside and outside the metaverse}
\label{t_threats}
\begin{tabular}{ccc}\toprule
\textbf{   }&     \textbf{Inside}&            \textbf{Outside}     \\ \hline

\textbf{Threats}    & \multicolumn{1}{m{4cm}}{deepfake in the real identity and information.}   &  \multicolumn{1}{m{4cm}}{deepfake in the virtual characters and virtual scenes.}    \\\hline

\textbf{Examples}    & \multicolumn{1}{m{4cm}}{Fake audio and video productions, property theft, political attacks, business defamation, \ldots}      &  \multicolumn{1}{m{4cm}}{Virtual identity impersonation, virtual contents manipulation, virtual social attacks, \ldots}   \\

\bottomrule
\end{tabular}
\end{table*}

\addtocounter{footnote}{-3}
\footnotetext[1]{\url{https://w.mgtv.com/b/512969/19015100.html}}
\addtocounter{footnote}{1}
\footnotetext{\url{https://ars.electronica.art/center/en/obama-deep-fake/}}
\addtocounter{footnote}{1}
\footnotetext{\url{http://innopixel.dk/projects/fredericia-display/index.html}}
\addtocounter{footnote}{1}
\footnotetext{\url{https://medicalfuturist.com/6-healthcare-examples-of-virtual-augmented-and-mixed-reality/}}

\subsubsection{Data Leakage}

Data leakage can occur in three phases: collection, transmission, and storage. First, in the metaverse, ubiquitous data such as facial expressions, eye/hand movements, speech, biological features, and brainwave patterns are collected during user activities, which the users may not be aware of. This kind of invasion of user privacy is quite common already before metaverse, e.g., Facebook and Cambridge Analytica scandal in 2018 resulted in the collection and use of data from millions of users without their consent~\cite{2018facebook}. Metaverse without strict regulations for data collection will likely feast much more user private data to the service providers. Second, private data can be leaked during data transmission and processing. For example, there are serious privacy risks when using voice interfaces with facial AR/VR devices. Shi et al. designed an eavesdropping attack called Face-Mic~\cite{shi2021facemic}, which utilizes subtle facial dynamics related to speech captured by the zero-permission motion sensors in AR/VR headphones to infer highly sensitive information, including the speaker's gender, identity, and speech content from real-time human speech. Lastly, storing sensitive user information in cloud servers or edge devices can also lead to privacy leaks. Hackers can use frequent queries through differential attacks to determine users' private information or launch DDoS attacks to compromise the entire cloud storage. Digital footprints, including avatar preferences, habits, and activities, can reflect the end users in the real world. Attackers can use these footprints to exploit real-world users. In the metaverse, where a wide third-person perspective is often used, users can also be tracked without their knowledge, and their user preferences can be used for social engineering attacks later on if they are not properly protected.

\subsubsection{Identity Security}

Biometric technology identifies a user based on their unique biological features, including fingerprinting, facial recognition, ear shape, eye scanning, lip movement, voice recognition, motion tracking, etc. Biometric technology is commonly used for user identification on phones, glasses, or browsers. There are serious risks associated with biometric authentication. One trend in biometric security is to use deep learning to generate fake biometric signals (such as facial images and videos) for biometric spoofing or presentation attacks, posing a direct threat to social security and personal privacy. In the metaverse, once a user's biometric data is stolen, the user's digital assets, avatars, social relationships, and digital life will be compromised in a more destructive way than the current Internet, because most biometric data cannot be reset. 

\subsubsection{Fraud Attack}


Deepfake technology can be used against AR/VR devices for disinformation attacks, identity theft to mislead the public, and financial fraud, etc. AR/VR devices consist of various data collection technologies, most of which collect sensitive information that is critical to the core functionalities. For example, VR-based spoofing attacks on facial recognition can use images from social networks to create users' facial models to attack facial authentication systems and spoof detectors using facial modes and animation for different facial expressions(e.g., smiling)~\cite{VR2016virtual}. In addition, AR technology can observe user behavior, which poses a threat to their privacy. For example, using computer vision and raw input data, AR applications can provide real-time attack guidance to an attacker's phone by mimicking a user's unique typing behavior on a smartphone without modifying the device, installing software, or using special hardware~\cite{wei2022all}. Alternatively, an attacker can perform a simulated attack by masquerading as another authorized entity to gain access to services or systems in the metaverse.

Indeed, fraud attack is a significant concern associated with deepfakes in the metaverse. The ability to create hyper-realistic synthetic media can enable malicious actors to deceive and manipulate others, leading to various fraudulent activities. The realistic nature of deepfakes poses a challenge for detection and increases the potential for fraudulent behavior within the metaverse environment.
With the continuous development and application of deepfake technology, we need to remain vigilant about these potential threats and take appropriate measures to prevent and address them.

\subsection{Countermeasures}
Some works have been proposed to detect deepfake content. 
For example, Sensity~\cite{sensity} is a detection platform similar to anti-virus software for deepfakes, which alerts users via email when they found synthetic media fingerprints generated by AI. Sensity employs the same deep learning process as creating fake videos. Zhu et al.~\cite{zhu2020blinkey} proposed ``blinkey'' to ensure user authentication on VR devices equipped with eye-tracking based on the rhythm of blinking that only the user knows. ``Blinkey'' claims to provide fast and accurate verification thanks to the unique genetic pattern that controls pupil dilation and constriction. 
Yang et al.~\cite{yang2020preventing} proposed a comprehensive lip feature representation and the extracted high-level lip feature has high discriminative power against human imposters, which can detect different kinds of deepFake attacks without any prior information. The proposed method is effective owing to its ability to capture the unique behavioral characteristics of individual speaker and the information asymmetry between the attacker and the authentication system.
Shen et al.~\cite{kim2022novel} proposed a voice spoofing defense system for AR headsets to defend against voice spoofing attacks. Aliman et al.~\cite{Aliman2020malicious} proposed a network security program to counteract deepfakes in VR. In addition, various methods have been developed to detect deepfakes, many of which rely on AI technology to detect facial swapping and other identity information forgery in images and videos.

To ensure the healthy development of the metaverse, it is necessary to be cautious about security risks and adopt a dual approach of legal management and technological measures to  promote a safe and secure metaverse ecosystem.

\section{FUTURE OUTLOOK}
\label{sec:five}

\subsection{Multimodal Authentication}
Metaverse encourages users to freely create personalized content and immersive experience. However, this can be easily exploited by deepfake attackers, leading to identity theft and impersonation risks. To address these potential risks and challenges, metaverse platforms need to enhance identity verification technology, such as adopting multimodal identity authentication, using various means to verify identities, and improving the accuracy and security of identity verification.


In the metaverse, multi-modal authentication can include facial recognition, voice recognition, biometric recognition, and smart contract. 
Facial recognition can be used for game character authentication, ensuring that only authorized users can access their virtual characters or virtual scenes. It can also be used for access management, such as restricting access to certain virtual scenes to specific users or groups of users. 
Voice recognition technology allows users to interact with the metaverse and authenticate their identity via voice. It can be used in scenarios such as voice interaction, audio, and video conferencing to ensure that only authorized users can perform relevant actions or participate in specific activities. 
Biometric recognition uses the user's biometric features, such as fingerprint, iris, and voice print, for authentication. It can be used in virtual scene access management to ensure that only authorized users can access restricted areas or perform sensitive operations. 
Smart contracts are automated contracts executed on the blockchain that can be used for operations such as trading digital assets and transferring virtual items. It can be used as a secure authentication mechanism to ensure that only authenticated users can carry out relevant transactions or operations.

Combining multiple identity verification methods can improve accuracy and security, leading to enhanced credibility and attractiveness of the metaverse. Therefore, multi-modal identity verification technology is of great significance for the development of the metaverse.

\subsection{Data Security Protection}
Without proper data protection measures, the misuse of deepfake technology may pose a great risk to the large amount of sensitive data collected in the metaverse, such as facial recognition and voiceprint recognition. Therefore, the metaverse needs to take a series of measures to strengthen the security of user data, to prevent misuse or malicious acquisition behaviors.

For example, the metaverse needs to adopt data encryption, access control, and permission management to ensure the security and confidentiality of user data. Multiple layers of security strategies should be established, and security monitoring mechanisms and preventative measures should be implemented. The metaverse needs to ensure the legal use and security protection of sensitive data. The metaverse also needs to adopt machine learning and other advanced technologies for real-time monitoring and identification of deepfake technology features.


\subsection{Deepfake for Complex Multi-object Scenes}

With the development of the metaverse, the forgery of complex scenes with multiple objects will be an important direction for the development of deepfake technology. In the future, deepfake technology may achieve more realistic multi-object scene forgery by using data collected from multiple cameras and sensors, as well as advancements in VR technology. Future deepfake technology and multi-object scene forgery techniques will require more complex algorithms and higher-resolution models to generate more realistic virtual scenes and characters. 

Multi-object scene forgery can be applied in numerous fields, such as creating complex VR experiences by blending real objects, scenes, and actions with virtual ones. This technology will help enhance the immersive experience of the metaverse, thus expanding people's virtual interaction and creative space. However, the misuse of deepfake and multi-object scene forgery technology may bring many ethical and legal issues, hence we need to strengthen the regulatory and management mechanisms of deepfake technology along the technology development.

\subsection{ChatGPT-enabled Deepfake Development}
Large language models like ChatGPT can provide more intelligent conversation interactions, enhancing users' virtual experience, which may have a positive impact on the development of VR and increase its application value in fields~\cite{ChatGPT-AR}, such as education, entertainment, and healthcare. ChatGPT can also be used to detect and identify textual information related to deepfake technology. For example, ChatGPT can be used for natural language understanding and sentiment analysis to help identify AI-generated  statements. 

On the other hand, text generated by ChatGPT can be used for speech synthesis, text generation, and generating fake social media accounts, among other applications, to create fake information or statements and deceive or mislead others. Overall, the impact of deepfake technology on society and individuals is complex, and large language models like ChatGPT may exacerbate its impact to some extent. If used correctly, deepfake technology and large language models like ChatGPT can also have a positive impact, promoting the development of VR and AI, and improving people's quality of life and experiences. Therefore, it is necessary to strengthen management and supervision when using these technologies to ensure their application can produce more positive effects.

\section{CONCLUSION}
\label{sec:six}
The rise of the metaverse brings more application scenarios and challenges for deepfake. This work surveys deepfake in the metaverse, including related technologies and application scenarios. It also summarizes the privacy and security concerns related to deepfake in the metaverse, as well as coping strategies. Finally, we give an outlook on the future development of deepfake in the metaverse. In short, with the development of the metaverse and the enhancement of deepfake technology, people can create their ideal world more freely, but the risks and challenges are inevitable. We hope that our proposal can ring a bell in this potential yet challenging futuristic direction.

\bibliographystyle{elsarticle-num} 
\bibliography{cas-refs}

\begin{thebibliography}{10}
\expandafter\ifx\csname url\endcsname\relax
  \def\url#1{\texttt{#1}}\fi
\expandafter\ifx\csname urlprefix\endcsname\relax\def\urlprefix{URL }\fi
\expandafter\ifx\csname href\endcsname\relax
  \def\href#1#2{#2} \def\path#1{#1}\fi

\bibitem{2017deepfake}
S.~Maddocks, ‘a deepfake porn plot intended to silence me’: exploring
  continuities between pornographic and ‘political’deep fakes, Porn Studies
  7~(4) (2020) 415--423.

\bibitem{S.3805}
S.3805 - malicious deep fake prohibition act of 2018,
  \url{https://www.congress.gov/bill/115th-congress/senate-bill/3805}.

\bibitem{tolosana2020deepfakes}
R.~Tolosana, R.~Vera-Rodriguez, J.~Fierrez, A.~Morales, J.~Ortega-Garcia,
  Deepfakes and beyond: A survey of face manipulation and fake detection,
  Information Fusion 64 (2020) 131--148.

\bibitem{lee2021all}
L.-H. Lee, T.~Braud, P.~Zhou, L.~Wang, D.~Xu, Z.~Lin, A.~Kumar, C.~Bermejo,
  P.~Hui, All one needs to know about metaverse: A complete survey on
  technological singularity, virtual ecosystem, and research agenda, arXiv
  preprint arXiv:2110.05352 (2021).

\bibitem{yadav2019deepfake}
D.~Yadav, S.~Salmani, Deepfake: A survey on facial forgery technique using
  generative adversarial network, in: 2019 International conference on
  intelligent computing and control systems (ICCS), IEEE, 2019, pp. 852--857.

\bibitem{zhang2022deepfake}
T.~Zhang, Deepfake generation and detection, a survey, Multimedia Tools and
  Applications 81~(5) (2022) 6259--6276.

\bibitem{4falchuk2018social}
B.~Falchuk, S.~Loeb, R.~Neff, The social metaverse: Battle for privacy, IEEE
  Technology and Society Magazine 37~(2) (2018) 52--61.

\bibitem{5liu2018cooperative}
H.~Liu, X.~Yao, T.~Yang, H.~Ning, Cooperative privacy preservation for wearable
  devices in hybrid computing-based smart health, IEEE Internet of Things
  Journal 6~(2) (2018) 1352--1362.

\bibitem{2023deepfake}
S.~Tariq, A.~Abuadbba, K.~Moore, Deepfake in the metaverse: Security
  implications for virtual gaming, meetings, and offices, arXiv preprint
  arXiv:2303.14612 (2023).

\bibitem{vasist2022deepfakes}
P.~N. Vasist, S.~Krishnan, Deepfakes: an integrative review of the literature
  and an agenda for future research, Communications of the Association for
  Information Systems 51~(1) (2022) 14.

\bibitem{mirsky2021creation}
Y.~Mirsky, W.~Lee, The creation and detection of deepfakes: A survey, ACM
  Computing Surveys (CSUR) 54~(1) (2021) 1--41.

\bibitem{7wang2022survey}
Y.~Wang, Z.~Su, N.~Zhang, R.~Xing, D.~Liu, T.~H. Luan, X.~Shen, A survey on
  metaverse: Fundamentals, security, and privacy, IEEE Communications Surveys
  \& Tutorials (2022).

\bibitem{8fernandez2022life}
C.~B. Fernandez, P.~Hui, Life, the metaverse and everything: An overview of
  privacy, ethics, and governance in metaverse, in: 2022 IEEE 42nd
  International Conference on Distributed Computing Systems Workshops (ICDCSW),
  IEEE, 2022, pp. 272--277.

\bibitem{9chen2022metaverse}
Z.~Chen, J.~Wu, W.~Gan, Z.~Qi, Metaverse security and privacy: An overview,
  arXiv preprint arXiv:2211.14948 (2022).

\bibitem{10zhao2022metaverse}
R.~Zhao, Y.~Zhang, Y.~Zhu, R.~Lan, Z.~Hua, Metaverse: Security and privacy
  concerns, arXiv preprint arXiv:2203.03854 (2022).

\bibitem{20203dfacegan}
S.~Moschoglou, S.~Ploumpis, M.~A. Nicolaou, A.~Papaioannou, S.~Zafeiriou,
  3dfacegan: Adversarial nets for 3d face representation, generation, and
  translation, International Journal of Computer Vision 128 (2020) 2534--2551.

\bibitem{MetaDeepFake}
The deepfake danger: When it wasn’t you on that zoom call,
  \url{https://www.csoonline.com/article/3674151/the-deepfake-danger-when-it-wasn-t-you-on-that-zoom-call.html}.

\bibitem{MetaverseDeepFake}
The metaverse is one giant deep fake,
  \url{https://www.linkedin.com/pulse/metaverse-one-giant-deep-fake-jerry-bui-cfe}.

\bibitem{15bose2019virtual}
A.~J. Bose, P.~Aarabi, Virtual fakes: Deepfakes for virtual reality, in: 2019
  IEEE 21st International Workshop on Multimedia Signal Processing (MMSP),
  IEEE, 2019, pp. 1--1.

\bibitem{2018facebook}
Facebook says nearly 50m users compromised in huge security breach,
  \url{https://www.theguardian.com/technology/2018/sep/28/facebook-50-million-user-accounts-security-berach}.

\bibitem{shi2021facemic}
C.~Shi, X.~Xu, T.~Zhang, P.~Walker, Y.~Wu, J.~Liu, N.~Saxena, Y.~Chen, J.~Yu,
  Face-mic: inferring live speech and speaker identity via subtle facial
  dynamics captured by ar/vr motion sensors, in: Proceedings of the 27th Annual
  International Conference on Mobile Computing and Networking, 2021, pp.
  478--490.

\bibitem{VR2016virtual}
Y.~Xu, T.~Price, J.-M. Frahm, F.~Monrose, Virtual u: Defeating face liveness
  detection by building virtual models from your public photos, in: USENIX
  security symposium, 2016, pp. 497--512.

\bibitem{wei2022all}
C.~Wei, W.~Lin, S.~Liang, M.~Chen, Y.~Zheng, X.~Liao, Z.~Chen, An all-in-one
  multifunctional touch sensor with carbon-based gradient resistance elements,
  Nano-micro letters 14~(1) (2022) 131.

\bibitem{sensity}
Sensity, \url{https://sensity.ai/deepfakes-detection/}.

\bibitem{zhu2020blinkey}
H.~Zhu, W.~Jin, M.~Xiao, S.~Murali, M.~Li, Blinkey: A two-factor user
  authentication method for virtual reality devices, Proceedings of the ACM on
  Interactive, Mobile, Wearable and Ubiquitous Technologies 4~(4) (2020) 1--29.

\bibitem{yang2020preventing}
C.-Z. Yang, J.~Ma, S.~Wang, A.~W.-C. Liew, Preventing deepfake attacks on
  speaker authentication by dynamic lip movement analysis, IEEE Transactions on
  Information Forensics and Security 16 (2020) 1841--1854.

\bibitem{kim2022novel}
J.-D. Kim, M.~Ko, J.-M. Chung, Novel analytical models for sybil attack
  detection in ipv6-based rpl wireless iot networks, in: 2022 IEEE
  International Conference on Consumer Electronics (ICCE), IEEE, 2022, pp.
  1--3.

\bibitem{Aliman2020malicious}
N.-M. Aliman, L.~Kester, Malicious design in aivr, falsehood and
  cybersecurity-oriented immersive defenses, in: 2020 IEEE International
  Conference on Artificial Intelligence and Virtual Reality (AIVR), IEEE, 2020,
  pp. 130--137.

\bibitem{ChatGPT-AR}
Y.~Ding, P.~Zhou, {Demo: Near Real-time ChatGPT-AR}, ACM MobiSys (2023).

\end{thebibliography}

\end{document}